\documentclass[12pt]{article}

\begin{document}
\title{{\bf Geometry of 2d spacetime and\\
  quantization of particle dynamics}}
  
 \author{George Jorjadze$^\dag$ and 
W{\l}odzimierz Piechocki$^\ddag$ \\
{\small $^\dag$Razmadze Mathematical Institute, 
Tbilisi, Georgia }\\
{\small $^\ddag$So{\l}tan Institute for Nuclear Studies,  
Warsaw, Poland }} 

\date{}
\maketitle

\begin{abstract}
\noindent
We analyze classical and quantum dynamics of a particle in 2d 
spacetimes
with constant curvature which are locally isometric but globally different.
We show that global symmetries of spacetime specify the symmetries of physical 
phase-space and the corresponding quantum theory. To quantize the systems
we parametrize the physical phase-space by canonical coordinates.
Canonical quantization leads to unitary irreducible representations
of $SO_\uparrow (2.1)$ group.

\end{abstract}

\section{Introduction}

Global properties of Lorentzian spacetime play an important 
role in the problem of unification of General Relativity and 
Quantum Mechanics. We present some results concerning the corresponding 
problem in the case of 2-dimensional spacetime with constant 
curvature $R_0\neq 0$. We consider classical and quantum dynamics of 
a relativistic particle in the following cases: 

(i) spacetime is a 
one-sheet hyperboloid ($R_0 <0$), 

(ii) spacetime is a stripe ($R_0 >0$),

(iii) spacetime is a half-plane ($R_0 <0$ and $R_0 >0$).

These examples of spacetime are  
(separately, for $R_0 <0$ and $R_0 >0$)
locally isometric  
but have different global properties. We have found the 
relations among the symmetries of
spacetime, physical phase-space
and corresponding quantum theory. For completeness 
of the present paper we recall some results of [1], where
we considered the dynamics of a particle in the Liouville
field to test 
our quantization method for gauge invariant theories. 
Some physical aspects of particle dynamics in the singular
Liouville field were discussed in [2]. 

\vspace{0.5cm}

Dynamics of a relativistic particle with mass $m_0$ in 
gravitational field $g_{\mu\nu}(x^0,x^1)$ is defined by the 
action
\begin{equation}
S=\int L(\tau )~d\tau ,~~~~L(\tau):=-m_0\sqrt{g_{\mu\nu}(x^0(\tau),
x^1(\tau))\;\dot{x}^\mu(\tau)\dot{x}^\nu(\tau)},
\end{equation}
where $\tau$ is an evolution parameter and $\dot{x}^\mu := 
dx^\mu/d\tau.$ The coordinate $x^0$ is associated with time and
it is assumed that $\dot x^0 >0$.

The action (1.1) is invariant under reparametrization $\tau
\rightarrow f(\tau).$ This gauge symmetry leads to the 
constrained dynamics in the Hamiltonian formulation [3].
The constraint reads
\begin{equation}
\Phi :=g^{\mu\nu}p_\mu p_\nu -m_0^2 =0,
\end{equation} 
where $p_\mu := \partial L/\partial\dot{x}^\mu$ are canonical 
momenta (we use units with $c=1=\hbar$). 

\vspace{0.5cm}

In what follows we assume that the 
physical phase-space is the set of
particle trajectories in spacetime [4,5]. This set can be identified 
with the space of gauge invariant dynamical integrals.
We parametrize this space by coordinates suitable for 
canonical quantization.

\setcounter{equation}{0}
\section{Geometry of 2d manifolds}

One can always choose local coordinates on 2d Lorentzian 
manifold in such a way that the metric tensor $g_{\mu\nu}$ 
takes the form [6]
\begin{equation}
g_{\mu\nu}(X)= \exp\varphi(X)\;\left( \begin{array}{cr}
  1&0\\0&-1 \end{array}\right),
\end{equation}
where $X:=(x^0,x^1)$ and $\varphi(X)$ is a field.

The scalar curvature calculated for (2.1) is given by 
\begin{equation}
R(X)=\exp\left(-\varphi(X)\right)(\partial^2_1 -
\partial^2_0 ) ~\varphi(X).
\end{equation}
Therefore, the case $R(X)=R_0= const$ leads to the equation for 
$\varphi$
\begin{equation}
(\partial^2_0 -\partial^2_1)\varphi (X) + R_0\exp\varphi (X)=0,
\end{equation}
which is known as the Liouville equation [7]. This equation is 
invariant under the conformal transformations of the metric (2.1)
\begin{eqnarray}
x^\pm \longrightarrow y^\pm(x^\pm),~~~~~~~~~~~~~~~~~~~~~~~~~~~~~~
\nonumber \\
\varphi (x^+,x^-)\longrightarrow \tilde{\varphi} (x^+,x^-):=
\varphi(y^+(x^+),y^-(x^-))+\log[{y^+}^\prime (x^+){y^-}^\prime 
(x^-)],
\end{eqnarray}
where $ x^\pm := x^0 \pm x^1$ and ${y^\pm}^\prime :=dy^\pm
/dx^\pm .$

The general solution to (2.3) is given by 
\begin{equation}
\varphi (x^+,x^-)=\log \frac{8{A_+}^\prime (x^+){A_-}^\prime 
(x^-)}
{|R_0| \left[ A_+(x^+)-\epsilon A_-(x^-)\right]^2},
\end{equation}
 where 
$ {A_\pm}^\prime :=dA_\pm/dx^\pm,~ \epsilon :=|R_0|/R_0.$

Solution (2.5) is invariant under the transformation (2.4) with
\begin{equation}
y^+(x^+) =A_+^{-1}\left (\frac{aA_+(x^+)+b}{cA_+(x^+)+d}\right ) ,~~~~
y^-(x^-) =A_-^{-1}\left (\frac{aA_-(x^-)+\epsilon b}
{\epsilon cA_-(x^-)+d}\right ),
\end{equation}
where $A_+^{-1}$ and $A_-^{-1}$ are the inverse of  $A_+$ and $A_-$
functions, respectively.

Taking $A_\pm(x^\pm)=x^\pm$ we get the solution
\begin{equation}
\varphi (x^+,x^-)=\log \frac{8}{|R_0|(x^+ -\epsilon x^-)^2}.
\end{equation}
 The conformal transformation (2.4) leads 
(2.7) to the general solution (2.5). Thus, we conclude that all 
Lorentzian 2d manifolds with the same constant curvatures are 
locally isometric. However, they can be globally different. 
Eqs.(2.1) and (2.3) do not define the 
global properties of spacetime manifold. In what follows we
consider the models of spacetime with different global properties
indicated in the Introduction.

\setcounter{equation}{0}
\section{Dynamics on hyperboloid}  

Let ($y^0,y^1,y^2$) be the standard coordinates on 3d Minkowski 
space with the metric tensor $\eta_{ab}=diag(+,-,-)$.
A one-sheet hyperboloid $\mathbf{H}$ is defined by
\begin{equation}
-(y^0)^2+(y^1)^2+(y^2)^2=m^{-2},
\end{equation}
where $m>0$ is a fixed parameter.

Making use of the parametrization
\begin{eqnarray}
y^0=-\frac{\cot m\rho}{m},~~~~y^1=\frac{\cos m\theta}{m\sin 
m\rho},~~~~y^2=\frac{\sin m\theta}{m\sin m\rho},\nonumber \\
{\mbox {where}}~~~~ \rho \in ]0,\pi/m[,~~~\theta \in [0,2\pi/m[ 
~~~~~~~~~~~~~~~~~~~~~
\end{eqnarray}
we get the induced metric tensor on $\mathbf{H}$
\begin{equation}
g_{\mu\nu}(\rho,\theta)=\frac{1}{\sin^2 m\rho}
\left( \begin{array}{cr}
1&0\\0&-1 \end{array} \right),
\end{equation}
which has the conformal form (2.1) with
\begin{equation}
\varphi(\rho,\theta)=-\log \sin^2 m\rho.
\end{equation}

Since $\mathbf{H}$ has a constant curvature $R=-2m^2~$ (see [6]), it is 
clear that (3.4) is the solution of (2.3) for $R_0 =-2m^2,$ 
where the parameters $\rho$ and $\theta$ are identified with 
the spacetime coordinates $x^0$ and $x^1,$ respectively.

The Lagrangian (1.1) in this case reads
\begin{equation}
L= - m_0~\sqrt{ \frac{\dot{\rho}^2-\dot{\theta}^2}
{\sin^2 m\rho}}~.
\end{equation}
We assume that trajectories are timelike $(|\dot{\rho}|>
|\dot{\theta}|)$ and $\dot{\rho}>0.$

The hyperboloid (3.1)  
is invariant under 
the Lorentz transformations, i.e., $SO_\uparrow (2.1)$ is the 
symmetry group of our system. The corresponding infinitesimal 
transformations (rotation and two boosts) are
\begin{eqnarray}
(\rho,\theta)\longrightarrow(\rho,\theta+\alpha_0/m),~~~~~~~~~~~~~~~~~~~~
\nonumber \\
(\rho,\theta)\longrightarrow(\rho-\alpha_1/m~ \sin m\rho\sin 
m\theta,\theta+\alpha_1/m~\cos m\rho\cos m\theta),\nonumber \\
(\rho,\theta)\longrightarrow(\rho+\alpha_2/m~\sin m\rho\cos m
\theta,\theta+\alpha_2/m~\cos m\rho\sin m\theta).
\end{eqnarray}
The dynamical integrals for (3.6) read
\begin{eqnarray}
J_0=\frac{p_{\theta}}{m},~~~~~J_1=-\frac{p_\rho}{m}\sin 
m\rho\sin 
m\theta +\frac{p_\theta}{m}\cos m\rho\cos m\theta, \nonumber \\
J_2=\frac{p_\rho}{m}\sin m\rho\cos m\theta 
+\frac{p_\theta}{m}\cos m\rho\sin m\theta,~~~~~~~~~~~~
\end{eqnarray}
where $p_\theta :=\partial L/\partial\dot{\theta},~p_\rho :=
\partial L/\partial\dot{\rho}$ are canonical momenta.

Since $J_0$ is connected with space translations (see (3.6)),
it defines particle momentum $p_{\theta}=mJ_0$.

It is clear that the dynamical integrals (3.7) satisfy the commutation
relations of $sl(2.\bf {R})$ algebra
\begin{equation}
\{ J_a , J_b \} =\varepsilon_{abc}\eta^{cd}J_d ,
\end{equation}
where $\eta^{cd}$ is the Minkowski metric tensor and 
$\varepsilon_{abc}$
is the anti-symmetric tensor with $\varepsilon_{012}=1$.

The mass shell condition (1.2) takes the form
\begin{equation}
\sin^2 m\rho~(p_\rho^2 -p_\theta^2) = m_0^2 ,
\end{equation}
which leads to the relation
\begin{equation}
J_0^2 - J_1^2 - J_2^2 = -a^2,~~~~~~~~~a:=\frac{m_0}{m}~.
\end{equation}
According to (3.1), (3.2) and (3.7) the trajectories satisfy
the equations
\begin{equation}
J_a y^a =0,~~~~~y_a y^a = - m^2,~~~~~
J_1y_2 - J_2y_1 = \frac{p_\rho}{m^2}~.
\end{equation}
Note that $p_\rho < 0$, since $\dot\rho >0$. Then, 
for a given point ($J_0, J_1, J_2$) of (3.10), Eq.(3.11)
uniquely defines the trajectory on the hyperboloid $\mathbf{H}$.
One can check that all points of the hyperboloid (3.10)
are available for the dynamics. Therefore, there is one-to-one
correspondence between the hyperboloid (3.10) and the space of
particles trajectories. 
Since the physical phase-space is assumed to be the space of 
trajectories [4,5], we conclude that the hyperboloid
(3.10) is the physical phase-space.

To quantize the system we introduce the cylindrical coordinates
$J \in {\bf R}$ and $\phi \in S^1$, 
which parametrize the hyperboloid (3.10) by
\begin{equation}
J_0 = J,~~~~J_1 = \sqrt{J^2 +a^2}\cos\phi ,~~~~
J_2 = \sqrt{J^2 +a^2}
\sin\phi  .
\end{equation}
The canonical commutation relation $\{J, \phi\}=1$ leads to (3.8).

In the `$\phi$-representation'
the set of functions $\psi_n= \exp {in\phi }$, 
 ($n\in Z$) form the basis of the Hilbert space $L_2(S^1)$.
For the corresponding quantum operators $\hat J_a$
we have to choose the definite operator ordering in (3.12). 
We specify it by the
following requirements:

a) the operators corresponding to (3.12) are self-adjoint, 

b) they generate global $SO_\uparrow (2.1)$ transformations, 

c) the quantum Casimir number is equal to the classical one, 
i.e., $\hat C =-a^2 \hat I$, where 
$\hat C :=\hat {J}_0^2 - \hat {J}_1^2 - \hat {J}_2^2$. 

These conditions can be satisfied by the operators
\begin{eqnarray}
\hat J_0 =\hat J = -i\partial_\phi ,~~~~
\hat J_+ = e^{i\phi}~
\sqrt{\hat J^2 +\hat J + a^2}, \nonumber \\
\hat J_- = \sqrt{\hat J^2 +\hat J + a^2}~
e^{-i\phi},~~~~~~~~~~~~~~~~~~~~~~~
\end{eqnarray}
where $\hat J_\pm :=\hat J_1 \pm i\hat J_2$ and we get
\begin{eqnarray}
\hat J_0 \psi_n = n\psi_n,~~~~\hat J_+ \psi_n =
\sqrt{n^2+n+a^2}~\psi_{n+1},
\nonumber \\
\hat J_- \psi_n =\sqrt{n^2 - n + a^2}~\psi_{n-1},~~~~~~
\hat C \psi_n =
-a^2\psi_n .
\end{eqnarray}

The case $a^2\geq 1/4$ corresponds to  
the unitary irreducible
representation (UIR) of the continuous principal series of 
 $SL(2.\bf {R})$ group, whereas the case $0< a^2<1/4$ presents 
the UIR of 
the additional continuous series [8].

Due to (3.14), momentum of a quantum particle 
$\hat p_{\theta}=m\hat J_0$ 
can take only discrete values
$P_n=mn, ~(n\in Z)$. This result is related to the fact that 
the space of the considered spacetime is compact.
 
\setcounter{equation}{0}
\section{Dynamics on half-plane ($R_0<0$)}

Let us consider the solution (2.7) for $\epsilon < 0$.
The Lagrangian (1.1) in this case reads
\begin{equation}
L = - 2a \sqrt {\frac{\dot x^+ \dot x^-}{(x^+ + x^-)^2}},
~~~~~~a :=\frac{m_0}{m},
\end{equation}
where $x^\pm :=x^0 \pm x^1$ and $m=\sqrt{-R_0/2}$. It is assumed that
$\dot x^\pm > 0$, which leads to $p_\pm < 0$.

Formally, (4.1) is invariant under the fractional-linear
transformations
\begin{equation}
x^+ \rightarrow \frac{ax^+ + b}{cx^+ + d}, ~~~~~
x^- \rightarrow \frac{ax^- - b}{-cx^- + d},~~~~~~ ad-bc =1.
\end{equation}
Thus, formally, $SL(2.\bf {R})/Z_2$ (which is isomorphic
to $SO_\uparrow (2.1)$) is the symmetry of our system.
But, the transformations (4.2) are well defined on the plane
only for $c=0$. The corresponding transformations (with
$c=0$) form
the group of dilatations and translations (along $x^1$),
which is a global symmetry of the spacetime.

The infinitesimal transformations for (4.2) are
\begin{equation}
x^+ \rightarrow x^\pm \pm \alpha_0,~~~~~~ 
x^+ \rightarrow x^\pm + \alpha_1 x^\pm,~~~~~~
x^+ \rightarrow x^\pm \pm \alpha_2 (x^\pm)^2
\end{equation}
and the corresponding dynamical integrals read
\begin{equation}
P = p_+ - p_-,~~~~~K = p_+x^+ + p_-x^-,~~~~M = p_+(x^+)^2
- p_-(x^-)^2 ,
\end{equation}
where $p_\pm =\partial L/\partial \dot x^\pm$.

The dynamical integrals (4.4)
satisfy again the commutation relations (3.8) with
\begin{equation}
J_0 =\frac{1}{2}(P + M), ~~~~~~
J_1 =\frac{1}{2}(P - M), ~~~~~~ J_2 = K .
\end{equation}
The constraint equation (1.2) in this case can be rewritten as
\begin{equation}
p_+p_-(x^+ + x^-)^2 = a^2,
\end{equation}
which leads to
\begin{equation}
K^2 -PM =a^2 .
\end{equation}

Eq.(4.4) defines particle trajectories on the plane
\begin{equation}
Px^+x^- + K(x^+-x^-)=M~.
\end{equation}

Eq.(4.1) shows that $x^0 =0$ is the singularity line 
in the spacetime. 
Particle needs infinite proper time to reach it [2].
This is why the dynamics of the particle can be considered for
$x^0 < 0$ and $x^0 >0$ separately.
In such interpretation we deal with two independent systems 
each without singularity [2].
Since the dynamics in both cases is similar, we consider
only the case $x^0>0$, i.e., 
spacetime is ${\bf {R}}_+\times {\bf {R}}$.

Since the half plane and the hyperboloid (3.1) have the same curvature
$R=-2m^2$, one can define the isometry map from two half-planes
$(\bf {R}_-\times \bf {R})\cup (\bf {R}_+\times\bf {R})$ to 
hyperboloid
\begin{equation}
y^0 =\frac{1-m^2x^+x^- }{m^2(x^++x^-)},~~~~~~
y^1 =\frac{1+m^2x^+x^- }{m^2(x^++x^-)},~~~~~~
y^2 =\frac{x^+-x^-}{m(x^++x^-)}~.
\end{equation}
 The map (4.9) is invertible 
and it covers almost
all hyperboloid except two generatrices given by $y^0 + y^1=0$.
Thus, the half-planes can be considered as two different patches of 
the hyperboloid (3.1).

According to (4.7), $K=\pm a$ for $P=0$. But since $p_\pm < 0$,
the trajectories with $P = 0$ and $K = a$ cannot exist for  $x^0 >0$
(see (4.4)). Any other 
point $(P,K,M)$ of the hyperboloid uniquely
specifies the particle trajectory.
Therefore, the physical phase-space is defined by the hyperboloid
(4.7) without the line $P = 0, K = a$. 
Hence,  $SO_\uparrow (2.1)$ is not the 
symmetry group of our classical system. The symmetry transformations
of the physical phase-space are translations and dilatations.
These transformations are generated by the dynamical integrals
$P$ and $K$, respectively. The transformations generated by $M$
are not defined globally.
Thus, the physical phase-space has the same symmetry as the spacetime.

To quantize the system we parametrize the physical phase-space
(which is isomorphic to $\bf {R}^2$) by
the coordinates $(p,q)$ [9] 
\begin{equation}
P=p,~~~~~~K = pq -a,~~~~~~~~M = pq^2 -2aq .
\end{equation} 
The canonical commutation relation $\{p,q\}=1$ provides
the commutation relations of $sl(2.\bf {R})$ algebra 
\begin{equation}
\{P, K\} =P, ~~~~~\{P, M\} =2K, ~~~~~\{K, M\} =M~. 
\end{equation} 
Applying the symmetric operator ordering in 
`q-representation' we get 
\begin{equation}
\hat P =-i\partial_q ,~~~~~
\hat K = -i q\partial_q -(a+\frac{i}{2}),~~~~~
\hat M= -i q^2\partial_q -2(a+\frac{i}{2})q,
\end{equation} 
which gives the realization of the classical 
commutation relations (4.11).

According to the quantization principles quantum 
observables should be represented by self-adjoint operators.
The operators $\hat P$ and $\hat K$ are self-adjoint and they
generate the group of translations and dilatations, 
which is the symmetry group of spacetime.
The symmetric ordering in `q-representation' leads to the 
symmetric
operator $\hat M$. The 
self-adjoint extension of $\hat M $ does exist, but it is 
not unique [1].
This ambiguity can be parametrized by continuous parameter 
$\alpha \in S^1$. For $\alpha =0$, we get UIR of 
$SO_\uparrow (2.1)$
group. The case $\alpha =\pi$ gives UIR of $SL(2.\bf {R})$. 
Other values of 
$\alpha$ lead to UIR of the universal covering group
$\widetilde {SL}(2.\bf {R})$. The Casimir number for all these 
representations is $C=-(a^2 + 1/4)$, 
but  these 
representations for different $\alpha$
are unitarily non-equivalent.

\setcounter{equation}{0}
\section{Dynamics on stripe\\ (and half-plane, $R_0>0$)}

Let us consider the spacetime to be a stripe
\begin{equation}
{\mathcal {S}}:=\{(t,x)~|~t\in {\bf {R}}, ~x\in] 0 ,\pi /m [ \},
\end{equation} 
with the metric tensor
\begin{equation}
g_{\mu\nu}(t, x)=\frac{1}{\sin^2 mx}
\left( \begin{array}{cr}
1&0\\0&-1 \end{array} \right).
\end{equation}
It defines the spacetime with constant positive curvature
$R=2m^2$.

The Lagrangian (1.1) in this case 
\begin{equation}
L= - m_0~\sqrt{ \frac{\dot{t}^2-\dot{x}^2}
{\sin^2 mx}}
\end{equation}
is invariant under the action of the universal covering
group $\widetilde {SL}(2.\bf {R})$.
The corresponding infinitesimal 
transformations  are
\begin{eqnarray}
(t,x)\longrightarrow(t-\alpha_0/m, x),~~~~~~~~~~~~~~~~~~~~
\nonumber \\
(t,x)\longrightarrow(t-\alpha_1/m~ \cos mx\cos mt,
x+\alpha_1/m~\sin mx\sin mt),\nonumber \\
(t,x)\longrightarrow(t-\alpha_2/m~\cos mx\sin mt,
x-\alpha_2/m~\sin mx\cos mt)~
\end{eqnarray}
and they lead to the dynamical integrals 
\begin{eqnarray}
J_0=-\frac{p_{t}}{m},~~~~~J_1=-\frac{p_t}{m}\cos 
mx\cos mt +\frac{p_x}{m}\sin mx\sin mt, \nonumber \\
J_2=-\frac{p_t}{m}\cos mx\sin mt 
-\frac{p_x}{m}\sin mx\cos mt,~~~~~~~~~~~~
\end{eqnarray}
which satisfy the commutation
relations (3.8).

The mass-shell condition (1.2)
\begin{equation}
\sin^2 mx~(p_t^2 -p_x^2) = m_0^2 ,
\end{equation}
leads to the relation
\begin{equation}
J_0^2 - J_1^2 - J_2^2 = a^2,~~~~~~~~~a:=\frac{m_0}{m}~,
\end{equation}
which defines two-sheet hyperboloid.

The physical condition $\dot{t}>0$ gives $p_t<0$. Thus,
the space
of the dynamical variables (5.5) is only the 
upper-hyperboloid ($J_0>0$).
According to (5.5) we have 
\begin{equation}
J_0 \cos mx = J_1\cos mt - J_2 \sin mt 
\end{equation}
and each point of the upper-hyperboloid defines the 
trajectory (5.8)
uniquely. These trajectories are periodic in time $t$ 
with the period $2\pi /m$. 
Hence, the space of trajectories 
has the symmetry of $SO_\uparrow (2.1)$ group, but not 
of $\widetilde {SL}(2.\bf {R})$.
Therefore, the upper-hyperboloid  describes the space of trajectories
and it can be considered
as the physical phase-space of the system.

Due to the translation invariance in time particle energy $E$
is conserved and from (5.5) we have $E=mJ_0$. Eq. (5.8)
shows that particle oscillates between the `edges' of space around the
stationary point $x=\pi /m$.

For the quantization we use the parametrization
\begin{equation}
J_0 = \frac{1}{2}(p^2 + q^2) +a,~~~J_1 = \frac{1}{2}p~\sqrt{p^2 + q^2 +4a},
~~~J_2 = \frac{1}{2}q~\sqrt{p^2 + q^2 +4a},
\end{equation}
where $(p,q)$ are coordinates on a plane. It is easy to see that (5.9) 
defines the unique parametrization of the upper-hyperboloid and the 
canonical commutation relation $\{p,q\}=1$ leads to (3.8).

To solve the ordering problem for the operators corresponding 
to (5.9), we again impose the requirements 
(see the case of one-sheet
hyperboloid):

a) the operators $\hat J_a$ corresponding to (5.9) 
are self-adjoint,

b) they generate $SO_\uparrow (2.1)$ global transformations,

c) the quantum Casimir number equals $a^2$.

Now, these requirements 
can be satisfied {\it {only}} for the {\it {discrete}}
 values of the 
parameter $a$
\begin{equation}
a =\sqrt {k(k-1)},~~~~~\mbox {with}~~~~~ k=2,3,4,....
\end{equation}
The corresponding operators $\hat J_a$ are
 \begin{equation}
\hat J_0 = a^+a^- +k,~~~~\hat J_+ = a^+\sqrt{a^+a^- +2k},
~~~~\hat J_- = \sqrt{a^+a^- +2k}~a^-,
\end{equation}
where $a^\pm :=(\hat p +i\hat q)/\sqrt 2$ are the 
creation and annihilation
operators, respectively.

The basis of the corresponding Hilbert space $\cal H$ is 
formed by the vectors of the Fock space $|n\rangle ~~( n\geq 0)$.

The spectrum for $\hat J_0$ reads
\begin{equation}
\hat J_0 |n\rangle=(n+k) |n\rangle
\end{equation}
and from (5.11) we get
\begin{equation}
\hat J_+|n\rangle = \sqrt{(n+1)(n+2k)}~|n+1\rangle,
~~~~\hat J_-|n\rangle = \sqrt{n(n+2k-1)}~|n-1\rangle
\end{equation}
Eqs (5.10)-(5.12) present the UIR of $SO_\uparrow (2.1)$ 
from the discrete
series of $SL(2.\bf {R})$ [8].

According to (5.12) the energy of a quantum particle $\hat E =m\hat J_0$
takes only discrete values $E_n=m(n+k)$, where $n$ is a nonnegative
integer.

\vspace{5mm}

Eq.(2.7) for $\epsilon >0$ defines the Liouville field on a plane
$$
\varphi (x^+,x^-) = -2\log m|x|
$$
with singularity at $x:=(x^+-x^-)/2=0$. The corresponding 
dynamics on the half-plane $x>0$ was considered in [1]. The global 
symmetry group of spacetime in this case is a group of dilatations
and translations along $t:=(x^++x^-)/2$ (similarly to the case
$R_0<0$). However, the space of trajectories has the symmetry
of $SO_\uparrow (2.1)$ group and the physical phase-space is identified 
with the upper-hyperboloid. Therefore, quantization in this case
can be done in the same way as for the stripe.

Due to the translation invariance in time, we again have conservation
of energy, but now its spectrum is continuous, since in this case
$E=m(J_0+J_1)/2$.

\vspace{5mm}

For the stripe (5.1) and the half-plane with $R_0>0$  
our quantization method leads to the result (5.10), i.e.,
for a fixed value
of space curvature $R_0=2m^2$,  particle mass $m_0$ can take {\it
only discrete} 
values given by
\begin{equation}
m_0=m\sqrt {k(k-1)},~~~~~~~~~~k=2,3,4,...
\end{equation}

\setcounter{equation}{0}
\section{Conclusion}

Two dimensional Lorenzian manifolds with constant curvature
are locally described by the Liouville field theory, which
has the symmetry associated with $sl(2.{\bf R})$ algebra (see (2.6)). 
Therefore,
the particle dynamics in the corresponding gravitational field
is characterized by three dynamical integrals. Due to the mass-shell
condition these integrals are on the hyperboloid (one-sheet hyperboloid
for $R_0 <0$ and upper-hyperboloid for $R_0>0$).
There is one-to-one correspondence between the space of available
dynamical integrals and the set of particle trajectories in spacetime.
This set of trajectories is the physical phase-space and it
depends on the global properties of spacetime.
Thus, the global properties of spacetime specify the physical phase-space
as an available part of the hyperboloid.

In the case when the space of particle trajectories has global
$SO_\uparrow (2.1)$ symmetry the physical phase-space is the entire
hyperboloid and our quantization method leads to the unique
quantum theory with $SO_\uparrow (2.1)$ symmetry.

We conclude: To quantize the system it is 
necessary to specify (or identify) the global properties of the spacetime.

 \vspace{0.5cm}

{\bf Acknowledgments   }


This work was supported by the grants from:
INTAS  (96-0482), RFBR (96-01-00344),
the Georgian Academy of Sciences and the
So{\l}tan Institute for Nuclear Studies.

\end{document}